# *Natural Language Processing in Health Professions Education: A Scoping Review*

Javad Mohammad Alizadeh
Barnett College of Public Health
*Temple University*
Philadelphia, USA
javad.m.alizadeh@temple.edu

Pegah Saebi
Tyler School of Art and Architecture
*Temple University*
Philadelphia, USA
pegah.saebi96@gmail.com

Mukesh Kumar Patel
Barnett College of Public Health
*Temple University*
Philadelphia, USA
mukesh.kumar.patel@temple.edu

Huanmei Wu
Barnett College of Public Health
*Temple University*
Philadelphia, USA
huanmei.wu@temple.edu

**Abstract**

Natural language processing (NLP) and artificial intelligence (AI) are rapidly transforming health professions education, yet no scoping review has systematically mapped their applications across the full spectrum of health education contexts, including public health. Following the Arksey and O'Malley framework and PRISMA-ScR guidelines, this review synthesized evidence from 64 studies published between 2015 and 2026, identified through searches of PubMed, ERIC, IEEE Xplore, and Google Scholar. Seven thematic domains were identified: automated assessment, large language models (LLMs) as student-facing learning support, virtual patients and clinical simulation, curriculum analysis and program evaluation, personalized and adaptive learning, public health and health promotion education, and educator and institutional integration. Findings reveal significant technical promise, particularly in automated assessment, clinical simulation, and curriculum analysis, alongside persistent challenges including hallucination and accuracy concerns, algorithmic bias, data privacy, digital divide, overreliance, and absence of standardized outcome measures. Only four studies explicitly addressed public health education, representing a major evidence gap. This review provides a foundation for evidence-informed integration of NLP and AI across health professions education and highlights public health education as a priority area for future research and investment.

## I. Introduction

The integration of AI and NLP into health professions education represents one of the most consequential pedagogical shifts of the current decade. NLP, a subfield of AI focused on enabling machines to process, understand, and generate human language, has evolved rapidly in recent years, from rule-based systems and classical machine learning (ML) approaches to transformer-based architectures and LLMs such as GPT-4 and its successors [1]. These advances have expanded the range of tasks NLP systems can support in educational contexts, including automated scoring of clinical documentation, generation of narrative feedback, development of conversational virtual patients, delivery of personalized learning recommendations, and large scale curriculum analysis [2].

Health professions education faces several distinctive challenges related to NLP. First, the volume of unstructured narrative data in medical training has long exceed educators' capacity for systematic analysis [2]. Second, competency-based medical education (CBME) frameworks demand continuous, high-frequency feedback that is difficult to sustain through manual process alone [3]. Third, the global health workforce crisis  with projected shortfalls of 10 million workers by 2030, particularly in low- and middle-income countries (LMICs), creates an urgent need for scalable education solutions [4].

The public release of ChatGPT in late 2022 markedly accelerated interest in AI applications across health education. Students, faculty, and institutions rapidly adopted these tools, often in the absence of formal pedagogical frameworks or governance structures [5]. Although prior reviews have examined AI in medical education broadly [6], the specific landscape of NLP applications across the full spectrum of health professions education, including public health, has not been systematically characterized. Public health education remains particularly underexplored. Its emphasis on population-level reasoning, health communication, and misinformation management presents great potential for NLP applications. However, the current evidence base for NLP in public health education remains limited [7].

In response, this scoping review systematically maps the existing literature on NLP and AI applications across health professions education, with attention to public health contexts. The objectives are to (i) characterize the approaches,

This research was partially supported by the National Science Foundation grant SaTC-2310298.

tools, and outcomes reported in the literature; (ii) identify cross-cutting challenges and evidence gaps; and (iii) provide recommendations to support responsible integration of NLP in health and public health education programs.

## II. METHODS

### A. Study Design

This study employed scoping review methodology following the framework proposed by Arksey and O'Malley, refined by Levac et al., while adhering to PRISMA-ScR reporting guidelines [8], [9]. A systematic search was conducted across PubMed, ERIC, IEEE Xplore, and Google Scholar in March 2026, with no restriction on publication date. The search combined terms related to NLP and AI methods (*e.g.*, natural language processing, text mining, large language models) with terms reflecting health education contexts (*e.g.*, medical education, public health, health informatics, health professions training) and educational processes (*e.g.*, curriculum, assessment, training, pedagogy). Search terms were applied at title level in PubMed and Google Scholar, and cross broader fields in ERIC and IEEE Xplore; the latter was additionally restricted to journal articles only.

### B. Selection Criteria

Studies were included if they met all three of the following criteria: (1) described the use of NLP, LLMs, or related AI-driven text-based methods; (2) were conducted within a health or public health education context (including medicine, nursing, allied health, health informatics, and health professions training); and (3) reported on methods, tools, integration approaches, or educational outcomes.

Studies were excluded if they met any of the following criteria: (1) used NLP solely as a research methodology without an educational application or learner population; (2) focused exclusively on clinical or administrative NLP without an educational component; or (3) were not available in English.

### C. Study Selection Process

The initial search retrieved 566 records (PubMed: 393, IEEE Xplore: 117, ERIC: 39, Google Scholar: 17). After removing duplicates, 553 unique records remained. Title screening excluded 432 records. Of the remaining 121 records, abstract screening excluded 48, leaving 73 articles for full-text review. Nine additional papers were excluded at full-text stage, resulting in a final sample of 64 studies. The study selection process is illustrated in Fig. 1.

### D. Data Extraction

Data were extracted using a structured form to ensure consistency and completeness across studies (see Table 1). The form captured the following elements: study title, authors, year of publication, publication venue, country of origin, study design and type, level of education, target learners, sample size, discipline, NLP/AI tool or method, integration approach, educational application, pedagogical framework, dataset or text source(*e.g.*, clinical notes, educational materials, learner-generated text), outcomes measured, key findings, relevance to public health, reported implementation barriers and facilitators, and study limitations identified by authors.

### E. Synthesis Approach

Given the heterogeneity in study designs, populations, and reported outcomes, a narrative synthesis approach was adopted. Following full-text review and structured data extraction, thematic analysis was applied inductively to the extracted data by two independent reviewers. Each reviewer independently read across the extracted fields. Reviewers then convened to compare their independently derived groupings and reached consensus through iterative discussion until all 64 studies were assigned to a domain. This inductive, data-driven process yielded thematic domains that reflect the predominant modes of NLP and AI application across health professions education.

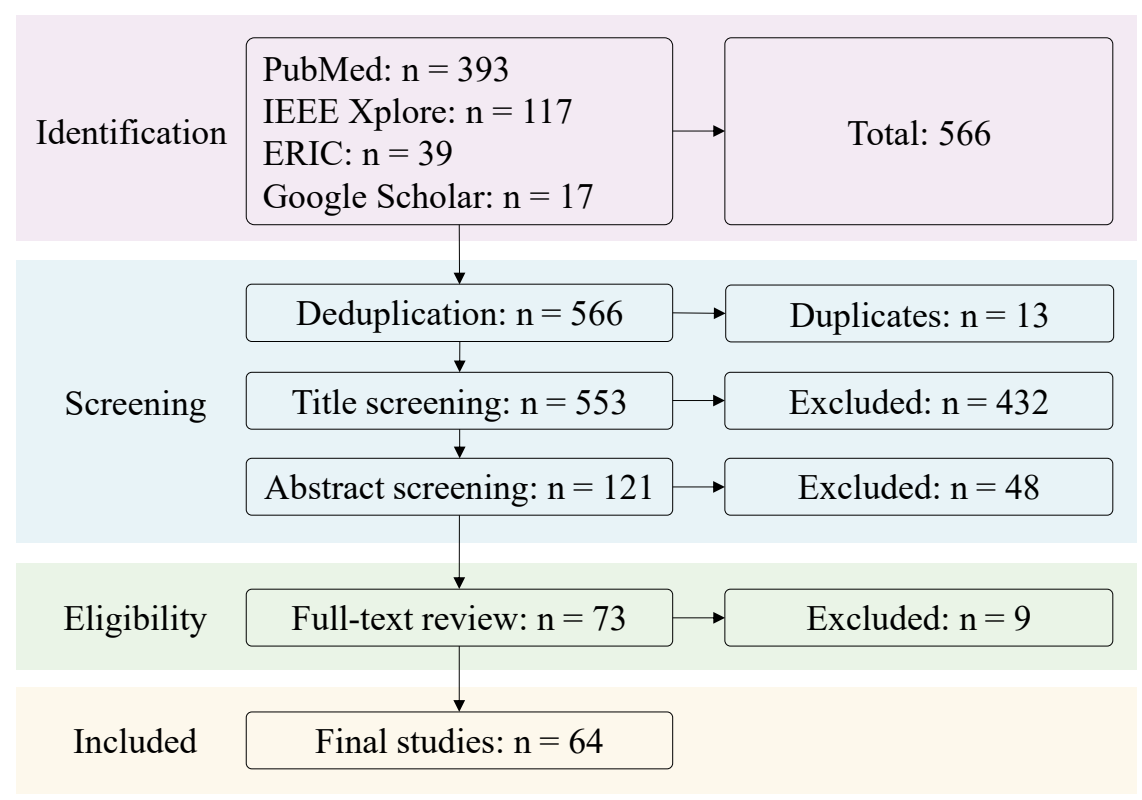

Fig. 1. Study selection process

Table 1. Data extraction form.

| Category | Information |
|---|---|
| Core reference | Title, Authors, Year, Journal/Source, Abstract |
| Study context | Country, Study Type, Level of Education, Target Learners, Sample Size, Discipline |
| NLP/AI details | NLP/AI Tool or Method, Integration Model, Educational Use of NLP/AI, Pedagogical Approach, Dataset or Text Source |
| Outcomes | Outcomes Measured, Key Findings, Public Health Focus, Barriers and Facilitators, Limitations |

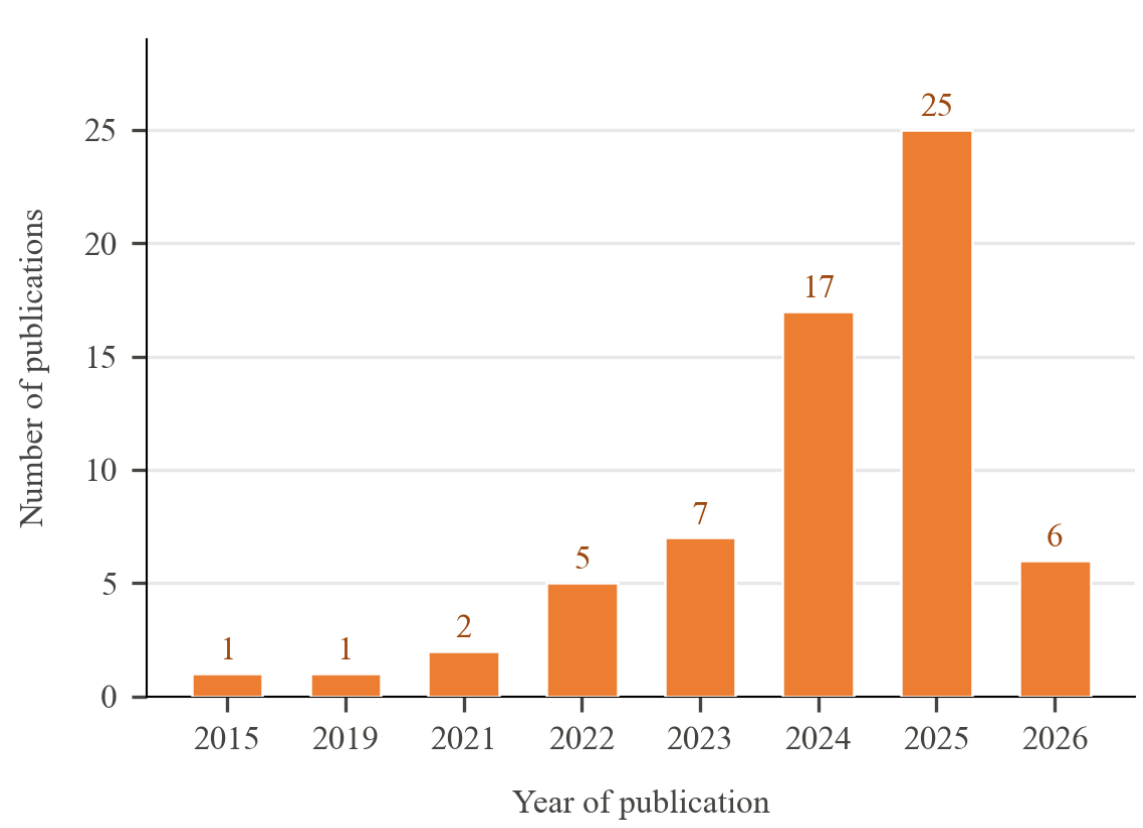

Fig. 2. Publication trend

Table 2. The number of literature for different categories.

| Category | Sub-types | # | % |
|---|---|---|---|
| Study Design | Empirical study | 37 | 58% |
| | Review (scoping, systematic, narrative) | 12 | 19% |
| | Commentary / Viewpoint / Guidance | 10 | 16% |
| | Mixed methods / Case study / Protocol | 5 | 8% |
| Target Learners | Undergraduate medical students | 21 | 33% |
| | Multiple / General / Not specified | 21 | 33% |
| | Postgraduate / Residents | 16 | 25% |
| | Allied health students | 5 | 8% |
| | Graduate students (non-clinical) | 2 | 3% |
| | Faculty | 2 | 3% |
| Disciplines | General Medicine | 32 | 50% |
| | Nursing & Allied Health | 9 | 14% |
| | Surgery | 6 | 9% |
| | Emergency Medicine | 6 | 9% |
| | Public Health | 4 | 6% |
| | Family Medicine / General Practice | 2 | 3% |
| | Radiology / Radiotherapy | 2 | 3% |
| | Anesthesiology | 1 | 1.6% |
| | Otolaryngology | 1 | 1.6% |

## III. Results

### A. *Overview of Included Studies*

The final 64 studies were published between 2015 and 2026, with yearly numbers shown in Fig. 2. Publication volume increased sharply in recent years, with 55 studies (86%) published between 2023 and 2026 and 25 studies (39%) appearing in 2025 alone. This trend reflects the rapid expansion of NLP and AI applications in health education following the emergence of LLMs. Notably, the earliest included study was published in 2015 [10], indicating that foundational NLP applications in health education predates the LLM era by nearly a decade. Study characteristics across design, learner type, and discipline are summarized in Table 2.

As shown in Fig. 3, title term analysis suggests a strong concentration of current research in in clinical training, assessment, and LLM applications. The word cloud visually summarizes the most frequent and prominent terms appearing across the included studies. The size of each word reflects how often it appears in the dataset where larger words indicate higher frequency and, therefore, greater emphasis in the literature. Furthermore, morphological variants were merged into a single root form (e.g., students and student, evaluations and evaluation), and standard stop words were removed prior to frequency counting.

### B. *Themes*

Seven thematic domains were identified from the included studies, spanning student-facing and institutional applications across health professions education (Fig. 4).

**Theme 1: NLP/AI for Student and Trainee Assessment**

Automated assessment was the most extensively studied theme, addressed across 22 studies spanning short-answer scoring, narrative workplace-based assessment, and clinical competency prediction. As shown in Table 3, structured text tasks consistently achieved near-human agreement, with proportion agreement reaching 0.97–0.99 and intraclass correlations of 0.893. Performance was more variable on complex narrative and multi-domain tasks, with BERT accuracy ranging from 41.6% to 92.5% depending on question type. Sentiment analysis tools produced mixed results across studies. A critical limitation was model failure to identify the bottom 10% of performers despite 80% overall accuracy, compounded by copy-paste behavior in approximately 49.6% of comments.

Sentiment analysis tools yielded mixed results: the National Research Council Canada (NRC) Emotion Lexicon analysis found specialty the sole predictor of positive word frequency in Entrustable professional activities (EPA) text [20], while the Valence Aware Dictionary and sEntiment Reasoner (VADER) found no significant differences in clerkship evaluation scores by student demographic [21].

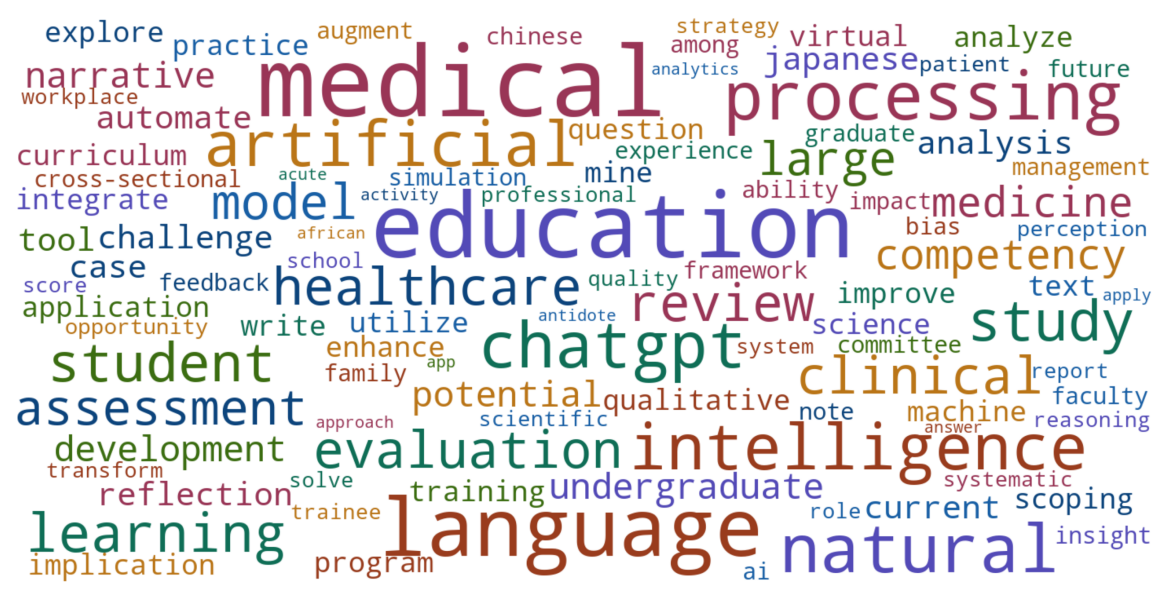

Fig. 3. The word cloud generated from the titles of the papers reviewed in this study.

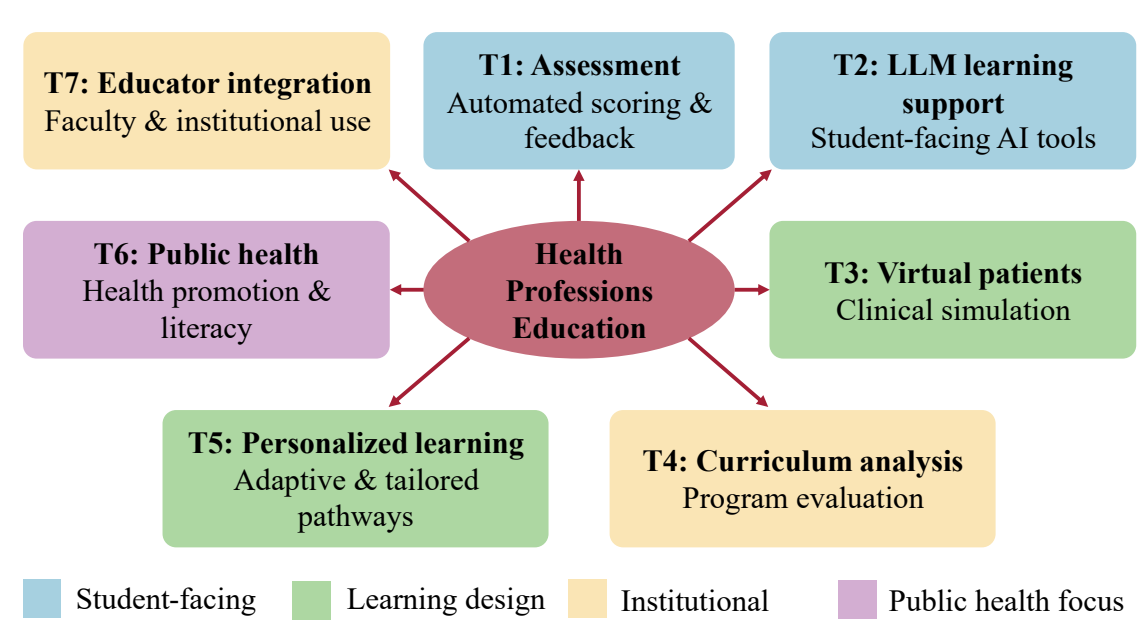

Fig. 4. Themes diagram

**Theme 2: LLMs as Student-Facing Learning Support**

LLM use for learning represented the second major theme across 19 studies spanning undergraduate medical, nursing, midwifery, physiotherapy, and health sciences students. LLM performance on health professions examinations showed variable but promising results: ChatGPT achieved 44%–64.4% accuracy on the United States Medical Licensing Examination (USMLE) Step 1 and Step 2, exceeding the passing threshold on the best dataset [25]; GPT-4 outperformed GPT-3.5 across licensing and board examinations (41.6%–84.7%) [26]; ChatGPT scored a median 4.0/5 on higher-order competency-based medical education biochemistry questions [27]; and 93.33% accuracy was reported on primary healthcare multiple choice questions (MCQs), supporting use as a self-directed resource in remote [28].

Student perception studies revealed consistent enthusiasm tempered by concern. Medical students used LLMs for concept clarification, MCQ preparation, and note-making while raising concerns about hallucination, privacy, and overreliance [29]. While 45.93% of undergraduates rarely used LLMs, attitude scores exceeded knowledge scores, suggesting openness without sufficient understanding [30]. In Uganda, 75.7% of medical students reported AI tool use, with urban and younger students significantly more likely to do so [31], and 92.59% of nuclear medicine students acknowledged ChatGPT's effectiveness with 69.44% reporting grade improvements [32].

Nursing, physiotherapy, and midwifery students preferred ChatGPT over traditional search engines but flagged accuracy concern [33], and midwifery students identified overreliance and reduced faculty interaction as risks [34]. LLM-based summarization in sleep medicine showed no significant difference compared to medical student performance [35], while a scoping review highlighted a geographic concentration in high-income countries and identified barriers such as the digital divide and language disparities [36].

Multiple reviews affirmed transformative potential for tutoring, quiz generation, and clinical reasoning while consistently flagging hallucination, academic dishonesty, and critical thinking erosion [37]-[40]. Real-world clinical ward use identified fabricated references as a key limitation [41].

**Theme 3: Virtual Patients and Clinical Simulation**

Ten studies addressed natural language processing (NLP) and artificial intelligence (AI) within virtual patient systems and clinical simulation environments, representing some of the most technically sophisticated applications in the corpus.

The AIteach system used Chinese NLP and word2vec to convert real medical records into question-and-answer training cases, with average test scores increasing from 69.87 to 85.6 among 15 graduate students ($p<0.01$), with logical thinking showing the most marked improvement (47%) [42]; an AI-powered clinical thinking training app extended this work across 479 students, with real-time feedback supporting more active self-directed learning [43].

The Hepius NLP-based virtual patient simulator, evaluated with fifth-year medical students across seven diagnostic reasoning dimensions including history taking, physical examination, and final diagnosis, revealed that 44% of students bypassed abdominal physical examination and proceeded directly to test ordering, exposing a methodological gap amenable to targeted intervention [44]. A hybrid systematic review confirmed that GPT-4o achieved 97.1% accuracy in data extraction comparable to human reviewers and that conversational virtual patients improve clinical competencies particularly in history-taking and clinical reasoning [45].

A multimodal simulation architecture further integrated automated speech-to-text transcription, Bidirectional Encoder Representations from Transformers (BERT)-based event detection, and physiological wearable data for performance assessment [24]. An LLM-based chatbot framework for radiotherapy education showed 70% of 60 pilot participants rated content helpfulness above average, with chatbot users achieving higher quiz scores [46]. Broader reviews identified Metaverse-based simulations with AI-generated symptom behaviors as a key emerging area [47], ChatGPT-4 demonstrated capacity to generate case-based simulation scenarios directly from educator prompts [48], and competency mapping accuracy of 92–97% for trainee documentation was reported across reviews [26], [49].

Table 3: Performance metrics for automated assessment studies (Theme 1).

| Ref. | NLP/AI Tool | Assessment Task | Metric | Value |
|---|---|---|---|---|
| [11] | ChatGPT 3.5 | Scoring medical student clinical notes | Scoring error rate (vs. standardized patients) | 1.0% vs. 7.2% (86% reduction) |
| [12] | Analysis of Clinical Text for Assessment (ACTA) | Short-answer questions, medical licensing exams | Proportion agreement with human scores | 0.97–0.99 |
| [13] | Automated grading system | Free-text responses, health informatics course | Intraclass correlation with human grading | 0.893; correction time ↓43% |
| [14] | BERT, logistic regression, random forest, XGBoost | Short answer marking, medical education | BERT classification accuracy | 41.6–92.5% (varies by question type) |
| [15] | Bag-of-n-grams NLP | Binary classification of emergency medicine (EM) resident narrative workplace-based assessment (WBA) comments | Accuracy | 87% |
| [16] | NLP + sentiment analysis | Surgical resident end-of-rotation text → Milestone ratings | Mean area under the curve (AUC) | 0.83 |
| [17] | BERT | Otolaryngology resident reflections / faculty feedback | Accuracy | 85% (reflections); 92% (feedback) |
| [18] | Fine-tuned BERT variants, FastText | Classifying anesthesiology feedback into Accreditation Council for Graduate Medical Education (ACGME) sub-competencies | Accuracy | Comparable performance reported; no specific value stated |
| [19] | Bio-ClinicalBERT | Predicting supervisor comment quality, emergency medicine | Accuracy | 87% |
| [22] | NLP classification model | Overall student performance stratification | Overall accuracy | 80%; failed to identify bottom 10% of performers |
| [23] | NLP sentiment analysis | Leniency bias correction in medical school narrative evaluations | Qualitative finding | Proof-of-concept; reliable distinction between high- and low-performers |
| [24] | Multimodal: BERT + speech-to-text + wearables | Simulation-based performance assessment | Architecture described | No single performance metric reported |

**Theme 4: Curriculum Analysis and Program Evaluation**

Thirteen studies applied NLP and text mining to analyze educational programs, evaluate curriculum quality, and identify patterns in student learning trajectories, demonstrating NLP's utility as an analytical lens on educational systems rather than as a direct instructional tool.

An early foundational application analyzed third-year medical student clinical notes to track geriatric competency exposure and trigger personalized email alerts, achieving 100% precision for advance directive identification [10]. A retrieval-augmented generation (RAG) NLP pipeline applied to 244,255 emergency department encounters across 62 residents quantified learning curves, finding the most rapid topic acquisition in postgraduate year 1 (PGY1) (42.1% of categories) with continued growth through PGY4 (63.2%), supporting the educational value of a four-year training model [50]. Custom phrase-matching NLP identified postoperative complications from surgical residency discharge summaries, detecting 68% of complication cases compared to 62% by traditional self-reporting, though the two methods captured largely non-overlapping case sets [51].

The TopEx NLP pipeline applied to 665 fourth-year medical student reflective writings identified 11 distinct topics in approximately two hours versus days for manual analysis, directly informing curriculum revisions [52]. Text mining of general practice/family medicine (GP/FM) residency reflections found strong co-occurrence between home visits and practice competencies [53], NLP analysis of student essays on AI in healthcare revealed underdeveloped patient-centered thinking [54], and co-occurrence network analysis of an AI ethics problem-based learning (PBL) program found significant knowledge gains and integration of previously fragmented thinking [55]. NLP methods including Latent Dirichlet Allocation (LDA) topic modeling, k-means clustering, and entity recognition have also been proposed for synthesizing literature on digital learning tools in postgraduate FM [56].

A scoping review of NLP in graduate medical education identified program evaluation, competency tracking, and learning analytics as key emerging applications [57] , and twelve practical tips for applying NLP to program evaluation framed the approach as scalable but methodologically complex [58]. A multi-agent LLM framework using Llama-3.1 and RoBERTa for analyzing open-ended resident feedback significantly outperformed single-agent LLMs in clarity (4.89 versus 3.00), semantic analysis, and coherence [59], and additional studies demonstrated NLP utility for detecting bias, tracking clinical exposure, and analyzing cross-cultural instructor experiences in international health education programs [2], [21], [60].

**Theme 5: Personalized and Adaptive Learning**

Ten studies addressed NLP and AI for delivering individualized learning experiences tailored to the competency gaps or learning trajectories of individual health professions students and trainees.

NLP-driven personalized email alerts based on clinical note content produced measurable behavioral changes in medical student documentation practices [10], and NLP mapping of trainee documentation to competency frameworks achieved 92%–97% accuracy, enabling adaptive curriculum design based on individual exposure gaps [49]. NLP-derived learning curve data revealed significant inter-resident variability in clinical topic acquisition, with a Gini coefficient indicating unequal exposure across trainees and providing an empirical basis for individualized remediation [50].

A comprehensive landscape review positioned adaptive learning platforms, predictive analytics, and NLP-driven competency tracking as the three most mature personalized learning technologies across undergraduate, graduate, and residency education [6], and NLP-based assessment analytics and intelligent tutoring systems were similarly identified as primary AI mechanisms supporting personalized learning [61]. NLP-based personalized learning recommendations in graduate medical education were further noted to reduce administrative workload while improving assessment accuracy [57].

A case study of AI and NLP integration in complementary medicine education highlighted the value of adaptive platforms for bridging linguistic and pedagogical gaps in multilingual, resource-limited contexts [62], and AI-driven adaptive learning environments within the Metaverse were identified as an emerging personalization frontier enabling simulation of complex scenarios tailored to individual learner profiles [47]. A review confirmed that NLP-enhanced assessment and feedback systems support automatic scoring and real-time feedback as a key pathway to personalized learning in health sciences education [63], and AI-enabled personalized learning and micro training in public health education were noted for their relevance to pandemic preparedness workforce development [7].

**Theme 6: Public Health and Health Promotion Education**

Four studies addressed public health education contexts, with an additional twelve reporting partial public health relevance, making this the most directly on-scope theme while also being the most underrepresented in the primary literature.

A scoping review of AI use in public health education for pandemic preparedness and response across 41 studies found conversational AI and algorithmic curation tools predominating, primarily supporting health literacy, risk communication, and misinformation management; interventions were concentrated in the response phase (52%), with fewer addressing preparedness (36%) and recovery (12%), and major evidence gaps included no standardized outcome measures, cost-effectiveness evaluations, and equity analyses [7].

An LLM-based multitier tagging system for Chinese online health education resources applied the Baichuan2-7B model with low-rank adaptation for automated metadata labeling of public health promotion content, achieving a Cohen $\kappa$ of 0.54 in agreement with human annotators, significantly higher than the human-human baseline of $\kappa$=0.32, identifying additional relevant tags missed by human annotators in 15.9% of cases with 90% expert-confirmed precision, and demonstrating scalable NLP infrastructure with a 94.8% processing success rate and median latency under one second [64].

Among studies with partial public health relevance, LLM performance on primary healthcare MCQs carried implications for health education in remote and underserved settings [28], a scoping review of ChatGPT in healthcare education identified limited LMIC representation and called for equitable global access [36], evidence from Uganda on widespread AI tool adoption among medical students highlighted implications for public health workforce training in low-resource settings [31]. ML and NLP applications in nursing education were noted for their relevance to community and primary care workforce development [65]. While NLP and AI are being applied in public health education particularly for health literacy, risk communication, and health promotion resource management, the evidence base remains thin, geographically concentrated, and methodologically heterogeneous, with a notable absence of standardized outcome measurement frameworks.

**Theme 7: Educator and Institutional Integration**

Eight studies focused on how faculty and institutions are integrating NLP and AI tools into professional workflows, curriculum design, and administrative operations, distinct from student-facing applications.

Table 4: Summary of the seven thematic domains.

| Theme | Primary NLP/AI Methods | Educational Function | Key Outcome(s) Reported | Refs |
|---|---|---|---|---|
| 1. NLP/AI for Student & Trainee Assessment | BERT, Bio-ClinicalBERT, FastText, bag-of-n-grams, ChatGPT, VADER, NRC Lexicon | Automated scoring of free-text responses; WBA narrative classification; competency milestone prediction | Scoring agreement 0.97–0.99 with human raters; AUC 0.83; 86% reduction in scoring error vs. standardized patients | [11]–[24] |
| 2. LLMs as Student-Facing Learning Support | ChatGPT (GPT-3.5 / GPT-4) | Exam preparation, tutoring, quiz generation, concept clarification | USMLE accuracy 44–64.4%; primary care MCQ accuracy 93.33%; 69.44% of nuclear medicine students reported grade improvements | [25]–[41] |
| 3. Virtual Patients & Clinical Simulation | Chinese NLP + word2vec, GPT-4o, BERT (speech-to-text + event detection), LLM chatbots | Interactive clinical reasoning training; simulation-based performance assessment | Test scores increased 69.87→85.6 ($p<0.01$); GPT-4o 97.1% accuracy in data extraction; 44% of students identified as bypassing physical exam | [24], [42]–[49] |
| 4. Curriculum Analysis & Program Evaluation | LDA topic modeling, k-means, RAG pipelines, TopEx, custom phrase-matching NLP, Llama-3.1 + RoBERTa multi-agent framework | Analysis of student reflections, residency notes, and learning trajectories to inform curriculum revision | 11 topics identified in ~2 hrs. vs. days manually; 100% precision for advance directive identification; multi-agent framework clarity score 4.89 vs. 3.00 (single agent) | [2], [10], [21], [50]–[60] |
| 5. Personalized & Adaptive Learning | NLP-driven alert systems, adaptive platforms, intelligent tutoring systems, competency-mapping NLP | Individualized feedback and learning recommendations based on competency gaps | Competency-to-framework mapping accuracy 92–97%; inter-resident variability quantified via Gini coefficient; reduced administrative workload | [6], [7], [10], [47], [49], [50], [57], [61]–[63] |
| 6. Public Health & Health Promotion Education | Conversational AI, algorithmic curation tools, Baichuan2-7B with low-rank adaptation | Health literacy, risk communication, misinformation management, metadata tagging of public health content | Cohen $\kappa = 0.54$ for automated tagging (vs. human-human $\kappa = 0.32$); pandemic response interventions 52%, preparedness 36%, recovery 12% | [7], [28], [31], [36], [64], [65] |
| 7. Educator & Institutional Integration | ChatGPT (MCQ/vignette/scenario generation), LLM chatbot frameworks, NLP curricula | Faculty workflow support; institutional governance; AI literacy training | 33% of faculty using ChatGPT for content generation; misinformation (74%) and plagiarism (65%) top faculty concerns | [6], [32], [46], [48], [66]–[69] |

A faculty survey at a Caribbean medical school found 33% using ChatGPT for MCQ and vignette generation, with primary concerns around misinformation (74%) and plagiarism (65%) [66], and three levels of integration in nuclear medicine education spanned student, faculty, and societal domains [32]. Integration of NLP and AI as formal curriculum at a Romanian medical school found ontologies and NLP to be the most challenging topics for students, leading to curriculum revision with expanded content on prompt engineering and hallucination detection [67]. A proposed framework for institutional deployment of LLM-based chatbots in radiotherapy education incorporated database management, quality control, monitoring and feedback loops, and ethical and legal safeguards [46].

A comprehensive institutional review described AI applications across five medical education domains, admissions, classroom learning, workplace-based learning, assessment, and program evaluation, identifying faculty development, AI literacy, and governance frameworks as the most pressing institutional needs [6]. AI-augmented competence committees in CBME were proposed to automate feedback summarization and bias detection while reducing committee workload, with emphasis on human oversight, bias auditing, and iterative validation before implementation [68]. Unguided student use of ChatGPT in speech-language pathology education produced problematic behaviors including ethical blind spots and overreliance, supporting the need for structured ethics integration and reflective practice embedded within curricula [69], and iterative prompt refinement through chain-of-thought processing was demonstrated to enable high-quality case-based simulation scenario development while reducing faculty workload [48]. Table 4 provides a summary of all seven thematic domains.

## IV. DISCUSSION

### A. *Summary of Key Findings*

This scoping review mapped 64 studies published between 2015 and 2026 examining NLP and AI integration across health professions education. While foundational NLP applications date to at least 2015 [10], 82.8% of included studies appeared between 2023 and 2026, reflecting a paradigm shift driven by the emergence of accessible LLMs. Seven thematic domains were identified: automated assessment, student-facing LLM learning support, virtual patients and simulation, curriculum analysis and program evaluation, personalized and adaptive learning, public health education, and educator and institutional integration. Across these themes, significant technical promise coexists with persistent methodological immaturity, equity gaps, and an underdeveloped evidence base for long-term educational outcomes. No single application domain has produced sufficiently large, multi-site, or controlled evidence to support widespread implementation recommendations, and the field remains dominated by single-institution studies, commentary, and proof-of-concept designs. The corpus spans a wide methodological range, from controlled empirical evaluations to commentaries and protocols, and findings should be interpreted with that heterogeneity in mind, as the evidential weight behind specific claims varies considerably across included studies.

### *B. Comparative Observations Across Domains*

The seven thematic domains identified in this review differ in their evidential maturity, methodological rigor, and readiness for implementation. Theme 1 represents the most developed application area, with 22 studies reporting quantitative performance metrics, including scoring agreement rates of 0.97–0.99 with human raters and intraclass correlations of 0.893, suggesting that NLP-based assessment of structured clinical text is approaching practical reliability for well-defined task types such as short-answer scoring and note evaluation. Theme 2 produced the largest volume of studies (19), driven largely by post-2022 interest in ChatGPT and GPT-4, yet evidence remains centered on examination performance proxies and student perception surveys rather than long-term learning outcomes.

Theme 3 demonstrate some of the most technically sophisticated implementations, combining NLP with intelligent tutoring components, and show promising pre-post outcome gains; however, sample sizes are small (as few as 15 students) and evaluation frameworks are not standardized across studies, making cross-study comparison difficult. Theme 4 stands apart methodologically: NLP here functions as an analytical lens rather than a direct instructional tool, and several applications achieved high precision (up to 100% for specific classification tasks), though these are largely proof-of-concept designs without benchmarking against human reviewers at scale. Theme 5 shows consistent accuracy in competency mapping (92–97%) but relies heavily on landscape reviews and secondary syntheses rather than primary experimental evidence, limiting conclusions about causal effectiveness.

Theme 6 is the most critically underdeveloped domain, with only four studies explicitly addressing this context and no standardized outcome measurement framework present across any of them. Finally, theme 7, with eight studies, is almost entirely descriptive, capturing adoption patterns and faculty attitudes, with no effectiveness data. Taken together, the evidence gradient across domains suggests that implementation efforts should prioritize areas with the strongest performance data (automated assessment, curriculum analysis), while treating remaining domains as active research priorities rather than implementation-ready applications.

### *C. Challenges*

**Accuracy, hallucination, and reliability.** The most frequently cited barrier was the risk of inaccurate, fabricated, or inconsistent outputs from LLMs and NLP systems. Hallucination was identified as a concern across student-facing LLM use [28], [37], [48], faculty-facing content generation [40], [66], and real-world clinical ward deployment where fabricated citations were documented [41]. For health education, where inaccurate information may directly influence clinical reasoning and patient care, hallucination represents a safety concern beyond academic integrity. NLP models trained on narrative assessment data are also subject to copy-paste artifacts and hidden linguistic codes that reduce model validity in ways that are difficult to detect without deep domain knowledge [22].

**Algorithmic bias and fairness.** Multiple studies identified the risk of NLP systems encoding and amplifying existing biases. Predictive models may obscure bias related to gender and ethnicity [16], and gender, racial, and age stereotypes embedded in NLP training corpora can propagate into outputs [58]. Sentiment analysis tools were found unable to distinguish meaningful qualitative differences from surface positivity in clerkship evaluations [21], and mandatory bias audits before deploying AI in competency committees have been called for [68]. These concerns are particularly acute in health professions education, where assessment fairness directly affects career trajectories and workforce diversity.

**Data privacy, security, and ethics.** The General Data Protection Regulation (GDPR) compliance was cited as a major institutional constraint [65], and concerns were raised about moving sensitive trainee data to external models for fine-tuning [18]. Legal and ethical safeguards were emphasized for any institutional chatbot deployment [46], and the absence of clear regulatory frameworks governing AI use in health education was noted across multiple studies [31], [39], particularly in low- and middle-income country contexts where institutional governance infrastructure may be limited.

**Digital divide and equity.** Research is overwhelmingly concentrated in the United States, China, and a small number of high-income countries [36]. Within a single country, medical students at rural Ugandan universities were 34% less likely to use AI tools than students at an urban university, reflecting infrastructure and awareness disparities [31]. The digital divide has been described as a fundamental barrier to AI integration in resource-limited education settings [62], and most AI-in-public-health-education interventions were developed in high- and middle-income settings, limiting applicability where the public health workforce need may be greatest [7]. Without proactive equity frameworks, AI risks exacerbating existing educational inequalities rather than democratizing access [49].

**Overreliance and academic integrity.** Overreliance was near-universal across student-facing LLM studies, with reduced engagement with traditional resources, diminished faculty interaction, and ethically problematic behaviors documented [28], [30], [33], [34], [69]. Academic dishonesty was a primary faculty concern [40], [66], collectively pointing to the need for governance frameworks rather than passive adoption.

**Lack of standardized outcome measurement.** The absence of standardized frameworks for measuring educational outcomes was identified as a critical methodological gap. No published studies assessed AI impact on critical thinking or clinical reasoning in undergraduate medical education [5], many studies lacked control groups or relied on self-assessment [26], and the absence of standardized learning outcome measures was explicitly flagged in public health education AI research [7]. This limits comparability across studies, prevents meta-analytic synthesis, and impedes translation of individual positive findings into scalable implementation evidence.

**Single-institution designs and generalizability.** Most empirical studies were single institution with small samples [15], [16], [18], [19], [22], limiting transferability of NLP models across different assessment cultures, languages, and health systems.

### *D. Recommendations*

Based on the synthesis of findings and challenges across 64 studies, this review offers the following recommendations for educators, institutions, policymakers, and researchers.

**Embed AI and NLP literacy into health professions curricula.** Formal AI literacy is needed as a core health education competency [5], [6], [49], [67]. Students need foundational understanding of NLP and LLM capabilities, limitations, and ethics, and practical curriculum integration including prompt engineering provides a concrete model [67], given widespread unguided student use [28], [30], [31]. A practical starting point is a standalone 2–3 hour module on LLM capabilities and limitations embedded within existing clinical reasoning or evidence-based practice courses, covering prompt construction, hallucination recognition, and ethical disclosure, competencies directly transferable to patient-facing contexts.

**Develop and adopt institutional governance frameworks.** Institutions must establish governance structures before deployment, covering data privacy, bias auditing, academic integrity, faculty training, and ethical disclosure [46], [58], [68]. Concretely, institutions can begin by adapting publicly available frameworks such as the UNESCO Recommendation on the Ethics of AI [70] or WHO's guidance on AI in health [71] as a governance template, then convening a standing faculty-student-IT committee with a mandate to review new AI tool adoptions before classroom deployment. LMIC institutions should be supported to develop context-specific frameworks rather than adopting high-income-country models.

**Prioritize equity in tool design and access.** Given consistent evidence of digital divide effects [7], [31], [36], [62], equity must be treated as a design principle rather than an afterthought. NLP tools for health education should be evaluated for accessibility across device types, language availability, bandwidth requirements, and cost. Parameter-efficient fine-tuning using low-rank adaptation (LoRA) can enable deployment of LLM-based health education systems on limited GPU infrastructure, offering a model for resource-conscious implementation [64]. Public health education programs serving rural, underserved, or low-resource populations require specific attention to ensure NLP and AI tools reduce rather than widen existing health workforce training gaps. In practice, this means requiring a brief accessibility audit, covering mobile compatibility, offline functionality, and language support, as a condition of institutional adoption, and piloting tools with students from underserved communities before broad rollout rather than as an afterthought.

**Standardize outcome measurement and invest in collaborative research.** Consensus on a minimum outcome set, covering learner outcomes, process outcomes, and contextual factors, is urgently needed, along with longitudinal data beyond immediate interventions. Funders and journals should require reporting against standardized frameworks. A minimum viable outcome set for NLP in health education could draw from existing frameworks such as Kirkpatrick's four-level training evaluation model, measuring reaction, learning, behavior, and results, and require at minimum a 90-day follow-up assessment to capture knowledge retention beyond immediate post-intervention testing. Multi-site collaborative studies would simultaneously address generalizability and equity gaps, producing evidence of broader validity than the pervasive single-institution literature currently allows.

**Direct specific research investment toward public health education.** The most consequential recommendation arising from this review is the need for targeted research on NLP and AI in public health education specifically, as distinct from clinical medicine. Proof-of-concept evidence exists for AI potential in public health literacy, risk communication, and health promotion resource management [7], [64]. However, applications in epidemiology education, health policy training, global health workforce development, and community health education remain almost entirely unexplored in the NLP literature. Funders such as HRSA, APHA education grants, and NIH's National Library of Medicine training programs represent concrete funding mechanisms through which investigators could initiate the first multi-site controlled studies of NLP integration in accredited public health curricula.

### *E. Implications for Public Health Education*

This review carries specific implications for public health educators. Public health's emphasis on population-level analysis, health communication, and misinformation management offers clear NLP application potential, yet structural differences from clinical medicine, including the absence of standardized patient-based curricula and lower reliance on narrative assessment, may explain its slower engagement with these tools.

NLP curriculum analysis tools and LLM-based health communication systems have direct transferability to public health field placements, health literacy training, and risk communication pedagogy [7], [52], [64]. Rapid AI adoption among health professions students in LMICs [31] underscores the urgency of equipping public health training programs with both the tools and governance frameworks.

### *F. Limitations of This Review*

Several methodological limitations apply. Search strategies were inconsistent across databases, as PubMed and Google Scholar used title-level terms while ERIC and IEEE Xplore applied broader field-level terms. IEEE Xplore was also restricted to journal articles only, excluding conference proceedings where emerging NLP and health informatics work is often first reported. The review covered only four databases, omitting Scopus, Web of Science, CINAHL, Embase, and ACM Digital Library; the absence of CINAHL in particular may explain the limited nursing and allied health representation in the corpus. No formal quality appraisal was conducted, consistent with scoping review methodology, meaning evidential weight varies considerably across included study types. Finally, restriction to English-language publications limits representation from non-English-speaking contexts, and given the rapid pace of LLM development, findings based on earlier model versions may already be outdated.

## V. CONCLUSION

NLP and AI demonstrate substantial potential across health professions education, but the evidence base remains dominated by single-institution designs, limited longitudinal outcome data, and a near-absence of equity-focused research. Public health education, the discipline most directly concerned with population health, health literacy, and workforce development for underserved communities, remains the least represented domain in the literature. Addressing this gap requires targeted research investment, interdisciplinary collaboration between health education and informatics communities, and governance frameworks ensuring responsible, equitable, and pedagogically grounded integration of NLP across the global health professions workforce.